\documentclass[aps, pra, superscriptaddress, amsmath, amssymb, reprint,floatfix,longbibliography]{revtex4-2}
\usepackage[english]{babel}
\usepackage{siunitx, mathtools, mathrsfs, bm, cancel, xfrac, nth} 
\usepackage[spaced=0]{diffcoeff}
\difdef{f, s, c, l}{}{op-symbol = d, op-order-nudge = 1 mu}
\usepackage[dvipsnames]{xcolor}
\usepackage{graphicx, circuitikz, tikz,  tkz-euclide, %subcaption
}
	\usetikzlibrary{calc, angles, positioning, intersections, quotes, decorations.markings}
	\usetikzlibrary{shapes, arrows, fadings}
\usepackage{pgfplots}
	\pgfplotsset{compat=newest}
\usepackage{bookmark}
\usepackage{hyperref}
\hypersetup{
	colorlinks=true,
	linkcolor=blue,
	breaklinks=true,
	filecolor=blue,      
	urlcolor=blue,
	citecolor=blue
}
\usepackage[capitalise]{cleveref}
\usepackage{ulem}

\renewcommand{\emph}[1]{\textit{#1}}
\newcommand{\ket}[1]{|{#1}\rangle}

\newcommand{\mK}{\: \mathrm{mK}}
\newcommand{\ohm}{\: \Omega}
\newcommand{\kohm}{\: \mathrm{k}\Omega}
\newcommand{\Hs}{\langle \hat{H}_S\rangle_{\!ss}}
\newcommand{\kb}{k_B}
\newcommand{\tr}{\mathrm{tr}}
\newcommand{\h}{{\sf h}}

\usepackage{amsmath}

\DeclareMathOperator*{\argmin}{argmin}

\begin{document}
\title{A thermometric quantum Brownian model for low-temperature electronics}

\author{Harry T. H. Fung}
\affiliation{School of Mathematics and Physics, The University of Queensland,  Brisbane, Queensland 4072, Australia}

\author{Thomas M. Stace}
\email{stace@physics.uq.edu.au}
\affiliation{School of Mathematics and Physics, The University of Queensland,  Brisbane, Queensland 4072, Australia}

\begin{abstract}
Quantum mechanical models of resistors in quantum electronics are often based on the quantum optical master equation (QOME). The QOME overlooks several fundamental properties, limiting its ability to model certain superconducting phenomena. Here we present a thermometric model of resistors, adapted from the quantum Brownian motion equation (QBME), that facilitates practical use of the QBME in modelling dissipative electronics at low temperatures. We compare the thermometric QBME presented here with predictions of the QOME, in the simple example of a transmon shunted by a resistor. We show that both the QBME and QOME yield comparable but physically distinct predictions, and discuss potential experimental tests with which to discriminate their use in modelling practical experiments.
\end{abstract}

\maketitle

\section{Introduction}

Superconducting quantum devices are often subject to resistive dissipation, such as external waveguides or other resistive shunts used to engineer desirable properties of the device \cite{cattaneo;2021,harrington;2022,nathan;2025}. Examples include DC SQUIDs and rapid single flux quantum logic circuits, which both use resistively shunted Josephson junctions (JJ) to eliminate hysteresis in their current-voltage characteristics \cite{fagaly;2006,tesche;1977,kumar;2019,likharev;1993}, and dissipatively error-corrected qubits for autonomous error correction \cite{touzard;2018,perez;2020,deNeeve;2022,sellem;2025,nathan;2025}.

A common approach to modelling dissipation in a quantised electronic circuit is based on the quantum optical master equation (QOME), which is of Lindblad form, \cite{gardiner;2004,breuer;2007,walls;2012,blais;2021,cattaneo;2021,rasmussen;2021,vaaranta;2022,hassler;2019}. Other quantum mechanical models for resistors also exist but are not as widely used in theoretical modelling. For example, \citet{nathan;2020} derive a Lindblad form that avoids the rotating-wave approximation used to derive the QOME, and apply this specifically to a resistor \cite{nathan;2025}. \citet{duffus;2016} treats a SQUID in an ohmic environment with a derivation similar to that of the quantum Brownian motion, though in the limit \(T \to 0\), by using a higher order expansion in the coupling operator and modifying dissipative terms to enforce a Lindbladian structure. 

These models have found practical utility for analysing dissipation. However, the physical properties of resistors impose additional constraints which are often neglected in practice. Firstly, in a circuit the voltage-flux identity \(V = \tfrac{d}{dt}{\Phi}\) and Ohm's constitutive law \(V = IR\) implies that resistive dissipation is invariant under translation of the magnetic flux \cite{murani;2020}. 

Secondly, the superconducting phase of a lumped superconducting circuit element \(\phi = \Phi/(\hbar/2e)\) may be either compact (\(\phi \in (-\pi, \pi]\) for JJs and capacitors) or noncompact (\(\phi \in \mathbb{R}\) for inductors). While the environment's effect on phase compactness is contested, for example \citet{murani;2020} assert that a resistive load need not grow the support of a compact phase coordinate of a system, while \citet{hakonen;2021} suggest that a resistive load necessarily `decompactifies' a compact coordinate, it is desirable to treat either topology when modelling superconducting devices \cite{liao;2025,murani;2020,hakonen;2021,murani;2021,devoret;2021,sonin;2022}.

Consequently, it is desirable to have a well-posed quantum mechanical model of a resistor that respects flux-translation invariance and is compatible with both compact and noncompact superconducting elements. However, as we discuss below, the QOME does not generally respect flux-translation invariance \cite{kohen;1997, coppola;2025,ghosh;2024}. This symmetry-breaking manifests in equations of motion (EOM) for first moments of system operators which disagree with the corresponding classical EOM, violating the correspondence principle (discussed in \cref{sec:background}). Consequently, the dynamics predicted by the QOME will necessarily differ from the classical predictions in regimes where they should agree. In addition, the EOM for the moments of the QOME imply  that the phase is noncompact.  

To attempt to resolve this, we revisit the the quantum Brownian motion master equation (QBME) because it is manifestly flux-translation invariant, making it a favourable foundation for modelling dissipative quantum circuits \cite{gardiner;2004,breuer;2007,kohen;1997,coppola;2025}. While the QBME formally respects flux-translation invariance, and the resulting EOM for first moments of system operators do agree with the corresponding classical EOM, it is not of Lindblad form. In particular, the Kossakowski form of the QBME superoperator has negative eigenvalues, making it explicitly non-Lindbladian 
\cite{massignan;2015,lampo;2016,lindblad;1976a}.

An \emph{ad hoc} fix to remove negative-eigenvalue Kossakowski-superoperators in the QBME is to add a `minimally invasive' term to cancel the troublesome eigenoperators (those with negative eigenvalue), thereby enforcing a Lindbladian form \cite{breuer;2007,christie;2024}. This guarantees the positivity of the density matrix dynamics. However, the additional terms are not physically motivated, and so the resulting EOM for second moments again differ from their corresponding classical counterparts. 

More generally, there is an inherent tradeoff in developing dissipative dynamical models: a master equation with terms that are quadratic in \(\hat{\Phi}\) and \(\hat{Q}\) (i.e.\ representing a dissipative harmonic system) can have at most two of the following three properties: (1) complete positivity, (2) flux-translation invariant dissipation, and (3) a thermodynamically consistent steady state  \cite{lindblad;1976b,kohen;1997,peixoto;1998,barchielli;2015}. It is likely that this tradeoff extends to nonlinear  systems as well. 

In this framing, the QOME ensures completely positive dynamics through its Lindbladian structure, and under a suitable rotating-wave approximation yields thermodynamic detailed balance, however it foregoes flux-translation invariance in the dissipation and therefore violates the correspondence principle at the level of the equations of motion. Conversely, the QBME relaxes detailed balance and does not guarantee completely positive dynamics, particularly at temperatures below the ground state energy, $k_B T<E_0$ \cite{gardiner;2004,homa;2019,lampo;2016}, but it does respect the correspondence principle in the equations of motion, including dissipative flux-translation invariance. 

Here, we introduce a modified QBME whose steady-state solution is thermometrically self-consistent at all temperatures. The resulting `thermometric' QBME is consistent with compact and noncompact phase topology, satisfies flux-translation invariance, and has a thermodynamically reasonable steady state solution, at the cost of being non-Lindblad. We apply this approach to model a transmon shunted by a resistor in two situations: first for the steady state of a driven system, and second for the dissipative evolution of an initially displaced nonlinear oscillator. 

 Using these example cases, we compare the predictions of  the thermometric QBME  with those of the QOME, finding that their steady state properties are similar, but with observable differences in their transient dynamics.

\section{Background}\label{sec:background}

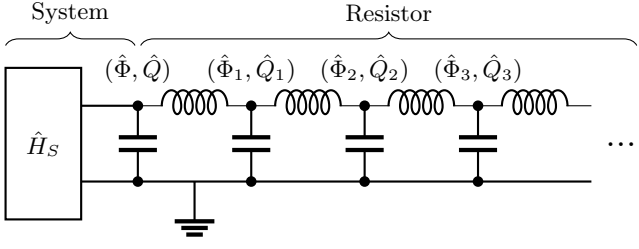
\begin{figure}[!]
    \centering
    \begin{circuitikz}[scale=1]
    \ctikzset{grounds/scale=1.25,capacitors/scale=0.6, inductors/scale=1}
    % \large
    % Some constants
    \def\l{2} % Length of system
    \def\w{1} % Width of system
    \def\fractn{1/4} % Fraction of \l to draw the height of the wires 
    \def\x{1.5} % Distance between nodes
    \def\N{3} % Number of nodes - 1 (ie N=3 gives 4 nodes)
    
    % Start of waveguide
    \draw[black, thick] (-\w, 0) rectangle (0, \l); % System
    \node at ({-\w/2}, {\l/2}) {\(\hat{H}_S\)};
    \filldraw[black, thick] (0, {\l*\fractn}) -- ({0.5*\x}, {\l*\fractn}); % Bottom
    \filldraw[black, thick] (0, {\l*(1-\fractn)}) -- ({0.5*\x}, {\l*(1-\fractn)}); % Top

    % Rest of the system
    \foreach \n in {0, 1, ..., \N} {
        \filldraw[black, thick] ({(\n+0.5)*\x}, {\l*\fractn}) -- ({(\n+1.5)*\x}, {\l*\fractn}); % Bottom wire
        \draw[black, thick]  ({(\n+0.5)*\x}, {\l*\fractn}) to [C, *-*] ({(\n+0.5)*\x}, {\l*(1-\fractn)}); % Capacitors
        \node[right=8pt] at ({(\n+0.5)*\x}, {(\l)/2}) {};
        \draw[black] ({(\n+0.5)*\x}, {\l*(1-\fractn)}) to[inductor] ({(\n+1.5)*\x}, {\l*(1-\fractn)}); % Inductors
        \node[below=2pt] at ({(\n+1)*\x}, {(\l)*3/4}) {};
        % \node[above=1pt] at ({(\n+0.5)*\x}, {\l*(1-\fractn+0.075)}) {{\((\hat{\Phi}_{\n}, \hat{Q}_{\n})\)}}; % Node labels
    }

    % Name the nodes of the system
    \node[above=1pt] at ({(0.5)*\x}, {\l*(1-\fractn+0.075)}) {{\((\hat{\Phi}, \hat{Q})\)}}; % Coupling node labels
    
    \foreach \n in {1, ..., \N} { % Rest of the nodes
        \node[above=1pt] at ({(\n+0.5)*\x}, {\l*(1-\fractn+0.075)}) {{\((\hat{\Phi}_{\n}, \hat{Q}_{\n})\)}}; % Node labels
    }

    \node[right=2pt] at ({(\N+1.5)*\x}, {\l/2}) {\Large ...}; % Ellipsis at the end of resistor
    \draw[black, thick] ({\x}, {0.49}) -- ({\x}, {0.5}) node[ground]{};
    % \draw[black, thick, ->] ({\x}, {\l*\fractn}) -- ({\x}, {0});
    % System curly bracket
    \draw[decorate,decoration={brace,amplitude=5pt,raise=4ex}]
    (-\w, {\l-0.25}) -- (0.475*\x, {\l-0.25}) node[midway,yshift=3em]{System};
    % Resistor curly bracket
    \draw[decorate,decoration={brace,amplitude=5pt,raise=4ex}]
    (0.525*\x, {\l-0.25}) -- ({(\N+1.9)*\x}, {\l-0.25}) node[midway,yshift=3em]{Resistor};
\end{circuitikz}
    \caption{The waveguide model of a resistor, consisting of an infinite series of capacitors and inductors, connected to a system. The leftmost nodes are the two resistor terminals.}
    \label{fig:circuit diagram}
\end{figure}

We begin from a standard approach to modelling resistive electronics by considering a system with internal Hamiltonian \(\hat{H}_S\) shunted by a resistor. We model the resistive element using Caldeira-Leggett theory as a semi-infinite waveguide \cite{caldeira;1983a,caldeira;1983b,vool;2017} (shown in \Cref{fig:circuit diagram}) that serves as an environmental bath with which the system interacts irreversibly. For our purposes, we consider galvanic coupling between the system and waveguide through a coupling node whose flux and charge operators are \(\hat{\Phi}\) and \(\hat{Q}\). This is analogous to coupling the position of a particle to the environment, which is typically how quantum Brownian motion is treated.

The Caldeira-Leggett model provides the foundation for deriving the QBME and QOME depending on the relative energy scale of the system, environment and coupling \cite{gardiner;2004,breuer;2007}.
A key assumption of the QBME is that the dynamical timescales in the system are slow compared to the relaxation timescale of the bath. Following standard derivations, the QBME for the resistively-shunted system is 
\begin{equation}
    \dot{\rho} = -\frac{i}{\hbar}[\hat{H}_S, \rho] - \frac{i \gamma}{2\hbar} [\hat{\Phi}, \{\dot{\hat{\Phi}}, \rho\}] - \frac{\gamma \kb T}{\hbar^2}[\hat{\Phi}, [\hat{\Phi}, \rho]], \label{eq:circuit qbm}
\end{equation}
where \(\rho\) is the system density matrix, \(\gamma= 1/R\) is the coupling strength, and $\dot{\hat{\Phi}}=\frac{i}{\hbar}[\hat H_S,\hat{\Phi}]$ \cite{gardiner;2004,vool;2017}. Due to its commutator structure, the $\gamma$-dependent dissipative terms in the QBME are manifestly invariant under translations in the flux coordinate, $\hat{\Phi}\rightarrow \hat{\Phi}+\Phi_c$ \cite{kohen;1997,coppola;2025}. It also makes no assumptions about the topology of the flux, thus being compatible with compact and noncompact descriptions of the flux or phase coordinate.

We note that the temperature dependence in \cref{eq:circuit qbm} is derived in the high-temperature limit \cite{gardiner;2004,breuer;2007,massignan;2015}. For the case of a harmonic oscillator, modifications to the temperature dependence of \cref{eq:circuit qbm} have been derived to describe low temperature dynamics \cite{caldeira;1989, hu;1992}. In what follows, we generalise the QBME, based on the principle that the temperature dependence in the QBME should be found self-consistently with the system itself.

A QOME can be derived from the Caldeira-Leggett model by assuming weak coupling to the bath and making a rotating-wave approximation, resulting in the well-known QOME for two-level or harmonic systems
\begin{equation}
    \dot{\rho} = -\frac{i}{\hbar}[\hat{H}_S, \rho] + \Gamma(\bar{n}_{\omega, T} + 1)\mathcal{D}[\hat{a}]\rho + \Gamma\bar{n}_{\omega, T}\mathcal{D}[\hat{a}^\dag]\rho, \label{eq:optical equation}
\end{equation}
where \(\Gamma\propto\gamma\) is the relaxation rate, $a^{(\dag)}$ is a system lowering (raising) operator between energy eigenstates, \mbox{\(\mathcal{D}[\hat{O}]\rho = \hat{O}\rho\hat{O}^\dag - \frac{1}{2}\{\hat{O}^\dag \hat{O}, \rho\}\)} is a dissipator, \mbox{\(\bar{n}_{\omega, T} = (e^{\hbar \omega/\kb T} - 1)^{-1}\)}, \(\omega\) is the transition frequency and \(T\) is the temperature of the environment \cite{gardiner;2004,breuer;2007,walls;2012}. Under additional approximations, \cref{eq:optical equation} can be  generalised to anharmonic systems with multiple non-degenerate transition frequencies. Importantly, the structure of the QOME ensures detailed balance, so that the steady-state solution of \cref{eq:optical equation} is the Gibbs state 
\begin{equation}
    \rho_G(T) = e^{-\hat{H}_S/\kb T}/Z, \label{eq:gibbs state}
\end{equation}
where $Z=\tr(e^{-\hat{H}_S/\kb T})$.

The difference between the QBME and the QOME can be demonstrated straightforwardly for a resistively shunted harmonic LC oscillator, whose system Hamiltonian is
\begin{equation}
    \hat{H}_\mathrm{LC} = \hat{Q}^2/(2C) + \hat{\Phi}^2/(2L),\label{eqn:LCham}
\end{equation}
where $\hat Q$ is conjugate to $\hat \Phi$, \(C\) is the capacitance, \(L\) is the inductance of the oscillator, and we define \(\Gamma = \gamma/C=1/RC\).

Using the QBME \cref{eq:circuit qbm}, the EOM for the first moments are 
\begin{align}
    \dot{\hat{\langle \Phi \rangle}} & = \langle \hat{Q} \rangle/C, \label{eq:qbme flux} \\
    \dot{\hat{\langle Q \rangle}} & = -\langle \hat{\Phi} \rangle/L -\gamma \langle \hat{Q} \rangle,
    \label{eq:qbme charge}
\end{align}
which agrees with the EOM for a classical RLC circuit \cite{caldeira;1983b}. Importantly, \cref{eq:qbme flux} is independent of $\langle\hat\Phi\rangle$ and $\gamma$, reflecting the fact that the dissipation is flux-translation-invariant.

Using the QOME \cref{eq:optical equation}, the corresponding EOM for the first moments are \cite{kohen;1997,coppola;2025}
\begin{align}
    \dot{\hat{\langle \Phi \rangle}} & = \langle \hat{Q}\rangle/C - \gamma\langle \hat{\Phi}\rangle/2, \label{eq:qome flux} \\
    \dot{\hat{\langle Q \rangle}} & = -\langle \hat{\Phi}\rangle/L - \gamma\langle \hat{Q}\rangle/2, \label{eq:qome charge}
\end{align}
which differs from those of the QBME and also the classical EOM, thereby violating the correspondence principle. In particular, since \cref{eq:qome flux} depends on $\langle\hat\Phi\rangle$, dissipation in the QOME explicitly breaks flux-translation invariance. Conceptually, this arises from the rotating-wave approximation, which neglects counter-rotating terms \cite{kohen;1997,ghosh;2024}, thereby importing the flux-translation dependence in $\hat H_S$ into the dissipative terms.

The functional advantages of the QBME -- flux-translation invariant dissipation, and the correct correspondence with the classical case -- make it an appealing foundational model for resistors. However, the steady-state solution to the QBME \cref{eq:circuit qbm} is typically only a valid quantum state when $\kb T$ is sufficiently large \cite{caldeira;1983a,homa;2019}.

This is again exemplified in an LC oscillator, for which the steady-state solution to the QBME satisfies 
\begin{equation}
    \Hs =\tr{(\hat H_S \bar\rho_{T})}= \kb T,\label{eqn:QBMEss1}
\end{equation}
where $\bar\rho_{T}$ is the steady-state solution to \cref{eq:circuit qbm} at temperature $T$ \cite{breuer;2007}. However, since the Hamiltonian is a bounded operator with $\hat H_S\geq E_0>0$ where $E_0$ is the ground state energy, any physical state $\rho$ (i.e.\ with $\tr{\rho}=1$ and $\rho\geq0$) should satisfy $\tr{(\hat H_S \rho)}>E_0$.  Thus, when $\kb T<E_0$,  \cref{eqn:QBMEss1} gives $\tr{(\hat H_S \bar\rho_{T})}<E_0$, which implies that the matrix $\bar\rho_{T}$ must have significant negativity, and is therefore unphysical.  That is, the steady-state solution $\bar\rho_{T}$ of the QBME in \cref{eq:circuit qbm} is manifestly unphysical when $\kb T<E_0$, and is therefore incapable of describing a physically-valid Gibbs-like thermal state at low temperatures. 
 
Instead, for $T=0$, the system should relax to a state close to the ground state of $\hat H_S$, and a physical model should yield \mbox{$\lim_{T\rightarrow0}\Hs=E_0$}. 
To address this, we introduce a thermometrically self-consistent modification to the QBME in \cref{eq:circuit qbm} that rectifies this limitation.

\section{The Thermometric QBME}

The key insight is to replace $k_B T$ in \cref{eq:circuit qbm} with a `thermometric parameter' \(\vartheta(T)\) that is a monotonic function of the temperature \cite{reif;1985}, resulting in a \emph{thermometric} QBME
\begin{equation}
    \dot{\rho} = -\frac{i}{\hbar}[\hat{H}_S, \rho] - \frac{i\gamma}{2\hbar} [\hat{\Phi}, \{\dot{\hat{\Phi}}, \rho\}] - \frac{\gamma\vartheta(T)}{\hbar^2}[\hat{\Phi}, [\hat{\Phi}, \rho]] \label{eq:generalised qbm}
\end{equation}
The preceding discussion requires that the thermometric parameter should satisfy 
\mbox{$\lim_{T\rightarrow0}\vartheta(T)\approx E_0$}, and \mbox{$\vartheta(T)\approx k_BT$} for $k_BT\gg E_0$, which we formalise below. This has some similarity with an `effective temperature' that has previously been used to describe the stationary state of the damped harmonic oscillator \cite{massignan;2015}. 

\begin{figure}[!]
    \centering
    \includegraphics{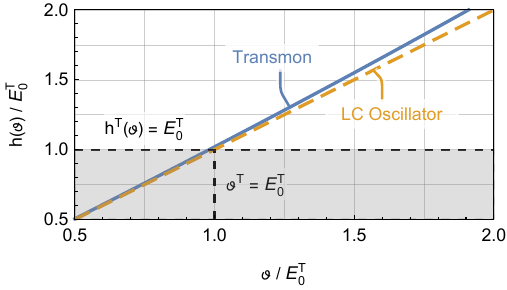}
    \caption{Steady state energy of a transmon \(\h^\mathrm{T}(T)\) (solid) and an LC oscillator \(\h^\mathrm{LC}(T)\) (dashed) as a function of the thermometric parameter \(\vartheta\). The shaded region below \(\h^\mathrm{T}(\vartheta) = E_0^\mathrm{T}\) indicates the unphysical region where \(\vartheta^\mathrm{T}\) does not correspond to a physical temperature. Device parameters used here and in subsequent figures are \(E_J/h = 25 \sqrt{2} \: \mathrm{GHz}\), \(E_C/h = (e^2/2C)/h = 1/\sqrt{8} \: \mathrm{GHz}\), \(E_L/h = \frac{1}{2L}\left( \Phi_0/2\pi\right)/h = 12.5 \sqrt{2} \: \mathrm{GHz}\). Calculated using the steady state of the thermometric QBME \cref{eq:generalised qbm}.}
    \label{fig:vartheta and H}
\end{figure}

At any given bath temperature $T$, we compute the thermometric parameter by constraining the steady-state energy expectation to match the  thermal Gibbs energy, which is derived from the  thermal Gibbs state of the system given in \cref{eq:gibbs state}.  Formally, for a given system, we define both the thermal Gibbs energy
\begin{equation}
E_G(T)\equiv\langle \hat H_S\rangle_G= \tr(\hat{H}_S \rho_G(T)),   \label{eqn:gibbs} 
\end{equation}
and the steady-state energy expectation function
\begin{equation}
    \h(\vartheta)\equiv \tr(\hat H_S \bar\rho_\vartheta), 
\end{equation}
 for all $\vartheta$ satisfying \mbox{\(\h(\vartheta) \geq E_0\)}, where $\bar\rho_\vartheta$ is the steady-state solution to \cref{eq:generalised qbm}. We then impose the thermometric self-consistency constraint $E_G( T)=\h(\vartheta)$, yielding an implicit relation between the thermometric parameter and the bath temperature
\begin{equation}
    \vartheta( T) = \h^{-1}(E_G(T)).\label{eqn:thermoimplicit}
\end{equation} 
For some systems these steps can be carried out analytically, but numerical solutions may be necessary in more complicated cases.

The energetic self-consistency condition is weaker than detailed balance, which is satisfied by the QOME, but it nonetheless provides a physically-motivated thermodynamic constraint on $\vartheta$, and is practically straightforward to implement for any given system. 
 
We now illustrate the application of the thermometric QBME in two cases: a harmonic LC oscillator, and an anharmonic transmon (or Cooper-pair box) system. For the LC oscillator, whose Hamiltonian is given in \cref{eqn:LCham}, it is straightforward to verify that 
\begin{equation}
    \h^{\mathrm{LC}}(\vartheta) = \vartheta,\label{eqn:hLC}
\end{equation} 
which is equivalent to \cref{eqn:QBMEss1} with the replacement $k_B T\rightarrow\vartheta$. 

A transmon (or Cooper-pair box) is a nonlinear system with Hamiltonian
\begin{equation}
        \hat{H}_\mathrm{T} = \hat{Q}^2/(2C) + E_J \cos(\hat\phi), \label{eq:transmon ham}
\end{equation}
where $\hat\phi=2\pi\hat\Phi/\Phi_0$. The spectrum of the transmon is anharmonic, and we compute \(\h^{\mathrm{T}}(\vartheta)\) numerically. 

\Cref{fig:vartheta and H} shows \(\h^{\mathrm{LC}}(\vartheta)\) and \(\h^{\mathrm{T}}(\vartheta)\). In both systems, the physically-relevant region is \(\h(\vartheta) \geq E_0\), and so we are concerned with values of $\vartheta$ for which $\h(\vartheta^{\mathrm{LC}}) \geq E_0^{\mathrm{LC}}$ for the LC oscillator, and similarly $\h(\vartheta^{\mathrm{T}}) \geq E_0^{\mathrm{T}}$ for the transmon.

\begin{figure}[t]
    \centering
    \includegraphics{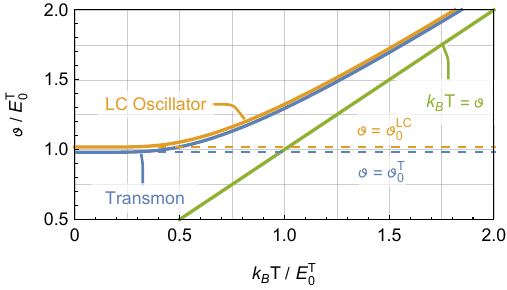}
    \caption{The thermometric parameter \(\vartheta\) as a function of the physical temperature \(T\) for the LC oscillator (see \cref{eq:thermometric lc}) and the transmon. Also shown is the thermometric parameter for the standard QBME \cref{eq:circuit qbm}, for which \(\vartheta = \kb T\). \(\vartheta_0^\mathrm{LC,T}\) is the value at which \(\h^\mathrm{LC,T}(\vartheta) = E_0^\mathrm{LC, T}\).}
    \label{fig:kT to vartheta}
\end{figure}

For the damped LC oscillator, the equilibrium Gibbs energy is
\begin{equation}
    E_G^{\mathrm{LC}}(T) = \hbar \omega_0\left(\langle \hat{a}^\dag \hat{a} \rangle_{G} + \tfrac12 \right) = \hbar \omega_0\left(\bar{n}_{\omega_0, T} + \tfrac12 \right),
\end{equation}
where \(\omega_0 = 1/\sqrt{LC}\) is the oscillator frequency \cite{vool;2017}. Using \cref{eqn:thermoimplicit,eqn:hLC}, we find that the thermometric parameter is given by
\begin{align}
    \vartheta^{\mathrm{LC}}( T) &=E_G^{\mathrm{LC}}(T), \nonumber\\
    &=E^{\mathrm{LC}}_0\coth\big(E^{\mathrm{LC}}_0/\kb T \big), \label{eq:thermometric lc}
\end{align}
where \(E^{\mathrm{LC}}_0 = \hbar\omega_0/2\). For such harmonic systems, the resulting master equation is identical to one derived by Caldeira and Leggett for a weakly-damped harmonic oscillator at arbitrary temperature \cite{caldeira;1989}. For an anharmonic transmon (or Cooper-pair box), the thermometric parameter is calculated numerically, using the same approach.

\Cref{fig:kT to vartheta} shows both $\vartheta^{\mathrm{LC}}( T)$ and $\vartheta^{\mathrm{T}}( T)$, normalised to the transmon ground state energy $E_0^\mathrm{T}$. Both satisfy the desired limits we previously outlined: \(\vartheta(T)\) approximates $E_0$ as $T\rightarrow0$, and asymptotes to \(\kb T\) at high temperature. This can be shown analytically for the LC oscillator by evaluating \cref{eq:thermometric lc} in the respective limits.

 The thermometric parameter is also equivalent to the effective temperature described previously by \citet{massignan;2015} for a harmonic system, although the master equation in \cite{massignan;2015} includes an additional `anomalous diffusion' term proportional to \([\hat{\Phi}, [\hat{Q}, \rho]]\) which comes from retaining additional terms in their derivation. The physical significance of this term in the context of resistors is unclear, as is the procedure for including it when considering an arbitrary system. Nonetheless, the consistency of the generalised thermometric QBME with prior results for linear systems suggests that it is a reasonable approach to extending the QBME to nonlinear devices at low temperatures.

 \begin{figure}[t]
    \centering
    \includegraphics{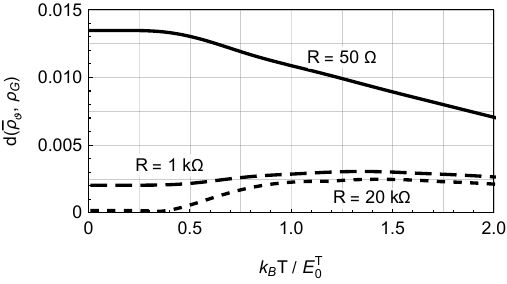}
    \caption{Trace distance between the QBME steady state \(\bar\rho_{\vartheta(T)}\) and the Gibbs state \(\rho_G\) \cref{eq:gibbs state} of a transmon as a function of temperature. Results are shown for resistances of \(50 \ohm\) (solid), \(1 \kohm\) (long dashes) and \(20 \kohm\) (short dashes).}
    \label{fig:trace distance}
\end{figure}

\begin{figure*}[t]
    \centering
    \includegraphics{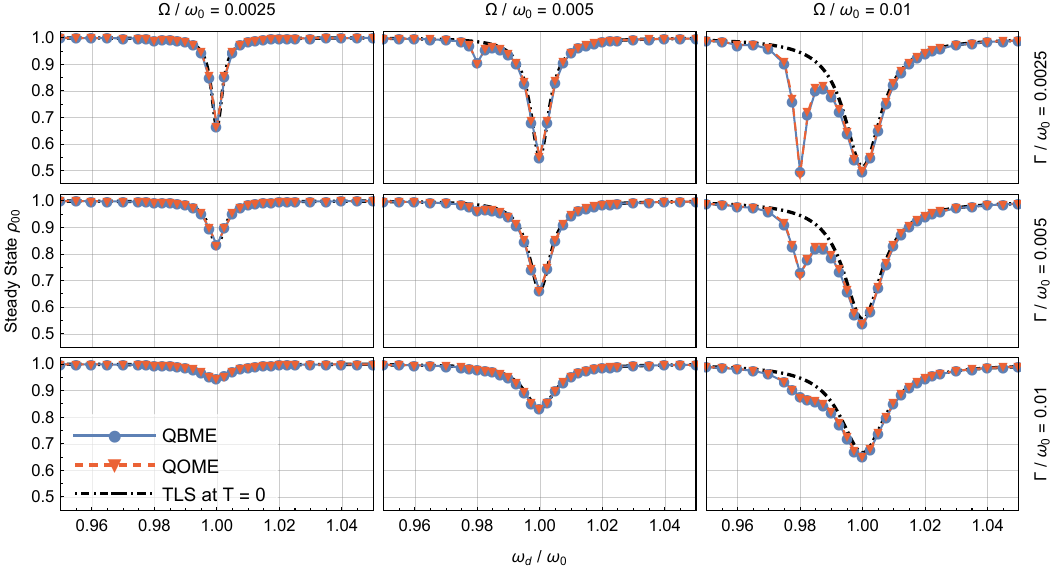}
    \caption{Each panel depicts the ground state population at equilibrium of a driven transmon shunted by a resistor (see \cref{eq:driven transmon ham}) as a function of the drive frequency. The dissipation rate increases down the rows (\(\widetilde{\Gamma} = \Gamma/\omega_0 = \{0.0025, 0.005, 0.01\}\)) and the Rabi frequency increases along the columns (\(\widetilde{\Omega} = \Omega/\omega_0 = \{0.0025, 0.005, 0.01\}\)). The ground state population is calculated with the thermometric QBME (circle, solid line), QOME (triangle, dashed line), and the analytical solution for a damped two-level system at zero temperature (dot-dashed line) (see \cref{eq:rho00 tls}). Temperature is \(T = 15 \mK\) (\(\kb T/\hbar\omega_0 \approx 0.03\)).}
    \label{fig:rho00 low T sweep}
\end{figure*}

While the thermometric QBME does not guarantee that the equilibrium steady state is the Gibbs state, we find in practice that the  thermometric constraint in \cref{eqn:thermoimplicit} leads to a steady state which is very close to the Gibbs state. This was previously shown analytically for the harmonic oscillator in the case where the anomalous diffusion term is neglected \cite{massignan;2015}.

We quantify the difference between the steady state of the thermometric QBME and the Gibbs state using the trace distance, 
\begin{equation}
   d(\bar\rho_{\vartheta(T)}, \rho_G) = \tr| \bar\rho_{\vartheta(T)} - \rho_G |/2,
\end{equation}
where \(| \rho | \equiv \sqrt{\rho^\dag \rho}\) \cite{neilsen;2010}. For the LC oscillator, we find $d(\bar\rho_{\vartheta(T)}, \rho_G) < 10^{-7} $. \Cref{fig:trace distance} shows the trace distance for a transmon, computed numerically. We see that it is small over all temperatures shown, and that the equilibrium thermometric QBME solution approaches the Gibbs state as the coupling strength decreases, as expected.

Since the Caldeira-Leggett model describes the environment as a bath of harmonic oscillators, we hypothesise that the true equilibrium state is some hybrid of a transmon and an LC oscillator Gibbs state, hence why our numerical results for \(\rho_{\vartheta(T)}\) coincide with \(\rho_G\) in the LC oscillator case. This is consistent with the idea that an open quantum system will, in general, thermalise to the `mean force' Gibbs state, which reduces to the Gibbs state in the limit of weak coupling \cite{lee;2022,trushechkin;2022,becker;2022,barnett;2024}. However, the similarity of the thermometric QBME steady state with the Gibbs state suggests that our chosen method of determining \(\vartheta\), that of energetic self-consistency with the Gibbs state, is suitable for a typical transmon.

\section{Numerical Thermometric QBME}

Having proposed the thermometric QBME, we numerically compare its predictions with those of the QOME, in two situations. The first case we consider is the steady state of a driven transmon shunted by a resistor. The second example is the dynamics of an initially displaced transmon. 

\subsection{Steady State of Driven Transmon}

\begin{figure*}[t]
    \centering
    \includegraphics{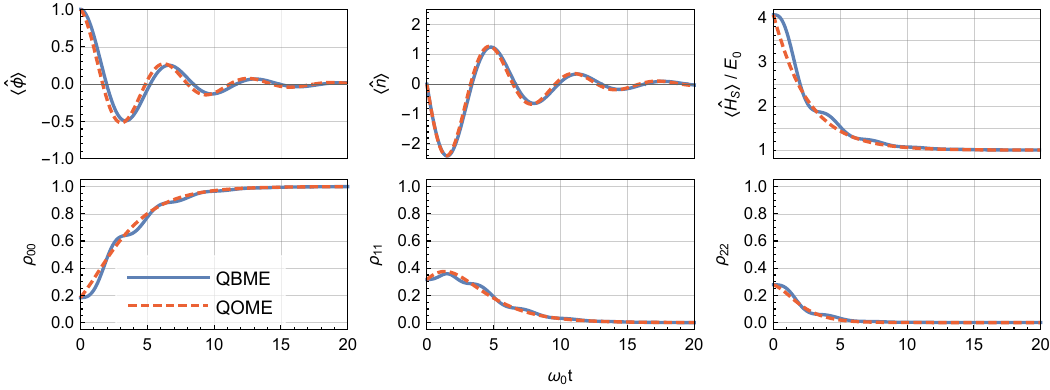}
    \caption{The expectation value of the phase, charge-number, energy, and the ground, first excited and second excited state populations of a shunted transmon as a function of time. The system was initialised in a flux-displaced state (\cref{eq:initial state}), with dissipation rate \(\widetilde{\Gamma} = 0.4\) and initial phase displacement \(\phi_c = 2\pi\Phi_c/\Phi_0 = 1\)). The evolution is evaluated with the thermometric QBME (solid) and the QOME (dashed) at \(T = 15 \mK\).}
    \label{fig:expectation values of displaced state}
\end{figure*}

We take the driven system to be governed by the Hamiltonian
\begin{equation}
    \hat{H} = \hat{H}_\mathrm{T} + \hat{H}_d = \hat{H}_\mathrm{T} + A\hat{\phi}\cos(\omega_d t), \label{eq:driven transmon ham}
\end{equation}
where \(\hat{H}_\mathrm{T}\) is the transmon Hamiltonian \cref{eq:transmon ham}, \(A\) is proportional to the Rabi frequency \(\Omega\), and \(\omega_d\) is the drive frequency.

Initialising the system in the ground state, we evolve the driven system to its steady state using the thermometric QBME and QOME. This is repeated while sweeping the drive frequency to build a profile of the steady state population of the ground state as a function of detuning. Several of these sweeps were evaluated for different values of \(\widetilde{\Omega} = \Omega/\omega_0\) and \(\widetilde{\Gamma} = \Gamma/\omega_0\), with a representative sample being presented in \Cref{fig:rho00 low T sweep}. In the figure, \(\widetilde{\Gamma}\) increases down the rows and \(\widetilde{\Omega}\) increases along the columns; the system is weakly driven in the bottom left panel, strongly driven at the top right panel and \(\widetilde{\Gamma} = \widetilde{\Omega}\) along the major diagonal. In each case, we have chosen a representatively small temperature, $T = 15\mK$, so that $\kb T/\hbar\omega_0 \approx 0.03\ll 1$.

In all panels, the steady-state solution for the ground-state population is very similar when calculated with the thermometric QBME and the QOME. In the  strongly driven regime, both solutions exhibit a resonance associated to the second excited state.

Since the temperature here is small, we also show the ground state occupancy for an equivalent two-level system at \(T = 0\), which is given by
\begin{equation}
    \rho_{00}^\mathrm{TLS}(\infty) = \frac{\Gamma^2 + 4\Delta^2 + \Omega^2}{\Gamma^2 + 4\Delta^2+2\Omega^2}, \label{eq:rho00 tls}
\end{equation}
where \(\Delta = \omega_d - \omega_0\) is the drive detuning \cite{walls;2012,breuer;2007}, which captures  the ground-to-first excited state transition in the panels shown. 

This example demonstrates that the steady state of the thermometric QBME is able to closely reproduce the predictions of the QOME across multiple drive strength and dissipation rates, at very low temperature. In addition, \Cref{fig:rho00 high T sweep} in \cref{app:high T sweep} presents the same sweep as in \Cref{fig:rho00 low T sweep} at a temperature where \(T \approx 0\) is no longer valid (\(\kb T/\hbar \omega_0 = 0.5\)), which shows the two models continue to agree very well when the temperature is comparable to the transition energy. These results establish the thermometric QBME as a practical and  accurate approach for studying steady-state properties.

\subsection{Dynamics from Displaced Ground State}

The second case we consider is the dynamical evolution of a transmon from an initially flux-displaced state. The transmon is initialised in a pure, displaced ground state wavefunction,
\begin{equation}
    \ket{\Psi_0} = \hat{\mathsf D}_{\hat{\Phi}}(\Phi_c)\ket{\psi_0},\label{eq:initial state}
\end{equation}
where \(\ket{\psi_0}\) is the ground state and $\hat{\mathsf D}_{\hat{\Phi}}(\Phi_c)$ is the flux displacement operator. In a laboratory setting, this experiment could be realised by rapidly switching a flux bias across a resistor-shunted transmon. 

\Cref{fig:expectation values of displaced state} presents the evolution of the expectation value of the superconducting phase, charge-number (\(\hat{n} = \hat{Q}/(2e)\)), energy, and the population of the ground, first excited and second excited state over the course of the decay. Broadly, the two models make comparable predictions. In both the QBME and QOME, the phase and charge-number exhibit decaying oscillations to zero, as expected of a weakly anharmonic system, the energy decays to \(E_0\), and the density matrix populations approach the ground state. However, a quantitative and qualitative difference is that the thermometric QBME predicts oscillatory decay in the energy and excited state populations, while the QOME predicts a simpler exponential decay that approximately coincides with the average value of the oscillations in the QBME dynamics. This has been previously noted for the LC oscillator when comparing the QOME with other models such as the standard QBME, Redfield equation and exact Hu-Paz-Zhang master equation \cite{kohen;1997,breuer;2024,farina;2019}. We also observe this behaviour when conducting the same simulation as in \Cref{fig:expectation values of displaced state} but for the LC oscillator.

The absence of oscillations in the system energy (and other quantities) under QOME evolution is due to the fact that the dissipation in the QOME couples symmetrically to both \(\hat{\Phi}\) and \(\hat{Q}\) (as in \cref{eq:qome flux,eq:qome charge}) \cite{kohen;1997}, which is distinct to the classical and QBME EOM where the dissipation couples only to $\hat{Q}\propto \hat{n}$ (as in \cref{eq:qbme charge}). Accordingly, we observe that the decay in the energy and excited populations is steepest when \(\langle \hat{n}\rangle\) is extremised, and decelerates to almost a stationary point when \(\langle \hat{n}\rangle = 0\) at which points the dissipation is essentially zero.

This example shows that there are distinct physical predictions from the thermometric QBME and the QOME during their dynamical evolution. Since the thermometric QBME predicts that the oscillations must become almost flat when \(\langle \hat{n}\rangle = 0\), the oscillation amplitude will be most significant when the decay is otherwise rapid. Conversely, if the overall dissipation rate is slow, the oscillations will have a small amplitude since there is little difference between the steepest points and the approximate stationary points.

\begin{figure}[!]
    \centering
    \includegraphics{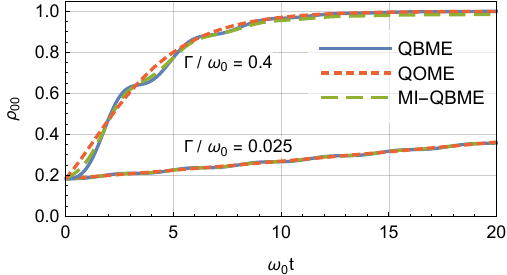}
    \caption{The ground state population of a shunted transmon during the decay from a flux-displaced state. The evolution is calculated using the thermometric QBME (solid), QOME (short dashed), and thermometric minimally invasive QBME (see \cref{eq:mi-qbme}) (long dashed) for two values of the dissipation rate and \(T = 15\mK\). The upper set of curves correspond to \(\widetilde{\Gamma} = 0.4\) (same as in \Cref{fig:expectation values of displaced state}) and the lower set correspond to \(\widetilde{\Gamma} = 0.025\).}
    \label{fig:displaced transmon ground state sweep}
\end{figure}

\Cref{fig:displaced transmon ground state sweep} compares the evolution of the ground state population for two different dissipation rates, \(\widetilde{\Gamma} = 0.4\) (same as \Cref{fig:expectation values of displaced state}) and \(\widetilde{\Gamma} = 0.025\). As above, the amplitude of the oscillations in the thermometric QBME's evolution increases with the dissipation rate, indicating that the predictions of the two models will be maximally distinct under conditions of high dissipation (i.e.\ small shunt resistance).

We reiterate that the absence of oscillations in the QOME evolution arises directly from the rotating-wave approximation, which neglects the counter-rotating terms.
These terms are negligible for \(\Gamma \ll \omega_0\), but start to become significant when \(\Gamma \sim \omega_0\), as seen in \Cref{fig:displaced transmon ground state sweep}. In this limit, the QBME offers a pathway to modelling the system dynamics in the intermediate damping regime, in this case around \(R \approx 750 \ohm\) for \(\widetilde{\Gamma} = 0.4\). At the very least, these results outline a key experimental signature that differentiates the dynamical predictions of the thermometric QBME and the QOME.

One of the main drawbacks of the QBME is that it does not guarantee complete positivity. However, we find that in practice, the thermometric QBME does not strongly violate positivity of the density matrix, leaving experimental predictions practically sound. 

\begin{figure}[!]
    \centering
    \includegraphics{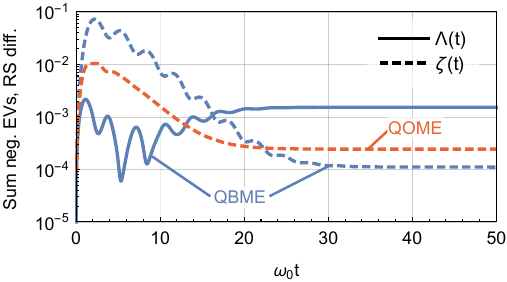}
    \caption{The sum of the negative eigenvalues of the density matrix \(\Lambda(t)\) (solid) and the Robertson-Schrodinger uncertainty relation \(\zeta(t)\) (see \cref{eq:RS uncertainty}) (dashed) for the thermometric QBME and QOME solution in \Cref{fig:expectation values of displaced state}. \mbox{\(\Lambda(t) = 0\)} for the QOME solution, hence it was omitted.}
    \label{fig:eigenvalues and RS uncertainty}
\end{figure}

The negativity of a density matrix can be quantified by calculating the sum of its negative eigenvalues, defined as \(\Lambda(t) = \sum |\lambda_{<0}|\), where \(\lambda_{<0}\) are the negative eigenvalues of \(\rho\). \Cref{fig:eigenvalues and RS uncertainty} shows \(\Lambda(t)\) for the thermometric QBME solution from \Cref{fig:expectation values of displaced state}, which shows that the sum of negative eigenvalues does not exceed \(3 \times10^{-3}\) and its value at equilibrium is \(\sim 1.5\times10^{-3}\). For the example with \(\widetilde{\Gamma} = 0.025\) in \Cref{fig:displaced transmon ground state sweep}, \(\Lambda(t)\) does not exceed \(4\times10^{-4}\ll1\). Rather than outright rejecting the solution due to this small negativity, we view it as a measure of physicality of the solution; small-negativity implies physically-reasonable model predictions. Indeed, it has been previously observed that the Redfield equation, which is also non-Lindblad, can have a greater accuracy than the QOME even when it has negative eigenvalues of a similar magnitude to those seen here \cite{hartmann;2020a,hartmann;2020b}. This includes correctly predicting oscillatory behaviour that was absent in a corresponding QOME \cite{hartmann;2020b}.

An indication that the thermometric QBME solution in \Cref{fig:expectation values of displaced state} is still physically reasonable is that its dynamical solution satisfies the Robertson-Schrodinger uncertainty relation for \(\hat{\Phi}\) and \(\hat{Q}\), which is given by
\begin{align}
    \zeta&\equiv\Delta \hat{\Phi}^2 \Delta \hat{Q}^2\!-\!\big(\langle \{ \hat{\Phi},\hat{Q}\} \rangle/2 \!-\! \langle \hat{\Phi} \rangle \langle \hat{Q}\rangle \big)^2\! -\big|\langle [\hat{\Phi}, \hat{Q}] \rangle /2\big|^2\nonumber\\
    &\geq 0, \label{eq:RS uncertainty}
\end{align}
where \(\Delta \hat{\mathcal{O}}^2 = \langle \hat{\mathcal{O}}^2 \rangle - \langle \hat{\mathcal{O}}\rangle^2\). This relation serves as a stronger version of the Heisenberg principle. A violation of \cref{eq:RS uncertainty} is a necessary but not sufficient condition for the density matrix to be negative, so it has been used as a method to detect positivity violations in solutions to the standard QBME \cite{homa;2019}. \Cref{fig:eigenvalues and RS uncertainty} depicts $\zeta(t)$ for both the thermometric QBME and the QOME solutions in \Cref{fig:expectation values of displaced state}. We conclude that solutions for which $\zeta(t)>0$ throughout the evolution remain physically reasonable, which is the case for the example shown, despite the small negativity in the QBME solution.

\section{Possible Extensions and Future Directions}

We comment here briefly about potential modifications to the thermometric approach described above.
Firstly, we assume here that the Gibbs energy is a good approximation for the thermodynamic equilibrium energy, which is consistent with the QOME. A more general description would use the mean force Gibbs energy, which self-consistently accounts for action of the system that modifies the environment \cite{lee;2022,trushechkin;2022,becker;2022,barnett;2024}. 

Secondly, alternative thermodynamic constraints could be imposed on $\vartheta$. For example, requiring that $\vartheta$ minimise the distance between $\bar\rho_\vartheta$ and $\rho_G(T)$,
\begin{equation}
    \vartheta_G(T)=\argmin_\vartheta ||\bar\rho_\vartheta-\rho_G(T)||,
\end{equation}
for a suitably chosen operator norm. This approach would adapt to more complex system-bath coupling terms, particularly in cases where there are multiple thermometric parameters $\{\vartheta_1(T),\vartheta_2(T),...\}$ appearing in an extended form of the QBME, \cref{eq:generalised qbm}. In practice, we have found this yields essentially the same results as shown here.

Lastly, we present an example of the thermometric parameter being applied to a variant of the QBME to demonstrate the flexibility of the procedure. Since we take flux-translation invariance to be necessary, another reasonable choice of master equation would be one that is completely positive but does not thermalise to the Gibbs state, as long as its steady state is adequately close to the Gibbs state. As we previously noted, a Lindblad variant of the QBME, which satisfies these properties, can be obtained by adding a `minimally invasive' term by hand \cite{breuer;2007}. Applying this procedure to the thermometric QBME gives what we call the thermometric MI-QBME
\begin{equation}
    \dot{\rho}_\mathrm{\text{MI-QBME}} = \dot{\rho}_\mathrm{QBME} - \frac{\gamma}{16\vartheta(T)}[\dot{\hat{\Phi}},[\dot{\hat{\Phi}}, \rho]], \label{eq:mi-qbme}
\end{equation}
where \(\dot{\rho}_\mathrm{QBME}\) is given in \cref{eq:generalised qbm}. The minimally invasive term involves only \(\dot{\hat{\Phi}}\), which makes it manifestly flux-translation invariant, and it can be straightforwardly verified that it is completely positive by calculating the eigenvalues of its Kossakowski matrix. For an LC oscillator, its steady-state energy function is
\begin{equation}
    \h^\mathrm{LC}(\vartheta) = \frac{(E_0^\mathrm{LC}/2)^2}{\vartheta} + \vartheta + \frac{\Gamma^2\hbar^2}{32\vartheta}, \label{eq:mi-qbme Hss}
\end{equation}
which can be determined by calculating the EOM of the second moments and solving for their steady state. \cref{eq:mi-qbme Hss} has a minimum with respect to \(\vartheta\) of 
\begin{equation}
    \h^\mathrm{LC}_\mathrm{min} = \sqrt{\left(E_0^\mathrm{LC}\right)^2 + \frac{\Gamma^2\hbar^2}{8}},
\end{equation}
which occurs at \(\vartheta = \h^\mathrm{LC}_\mathrm{min}/2\). The thermometric MI-QBME is clearly unable to reach the ground state energy for any finite coupling, which could pose an issue for its application to low temperatures if the dissipation rate is too large. However, for the dissipation rate we previously used (\(\widetilde{\Gamma} = 0.4\)), the minimum energy of \( \h^\mathrm{LC}_\mathrm{min} \approx1.04E_0^\mathrm{LC}\) makes it a potentially viable alternative to the thermometric QBME in the examples we have presented here.

An example of its use is depicted in \Cref{fig:displaced transmon ground state sweep} alongside the thermometric QBME and QOME, which shows that the thermometric MI-QBME coincides with the other solutions for \(\widetilde{\Gamma} = 0.025\) and also exhibits visible oscillations for \(\widetilde{\Gamma} = 0.4\). The amplitude of oscillations are smaller than those of the thermometric QBME, but their presence suggests that they are a physical phenomenon and are not simply due to the thermometric QBME lacking complete positivity. The key drawback of the thermometric MI-QBME is also evident in the figure, as the ground state population tends to a smaller value than the other two solutions, although in this case the discrepancy is minor.

Due to the \emph{ad hoc} nature of the minimally invasive term, there is no physical basis to believe that the MI-QBME is more accurate in general than the QBME and we leave a more thorough analysis of its effectiveness to future work. Nonetheless, this discussion shows that  thermometric self-consistency can be applied to models other than the QBME. 

\section{Conclusion}

We have presented a thermometric QBME with which to model resistors in superconducting devices.  This approach extends the QBME to handle low temperatures, respects flux-translation invariant dissipation, and is compatible with compact and noncompact superconducting phase descriptions. We used the thermometric QBME to model the evolution of a transmon shunted by a resistor, where we found that its steady state properties are quantitatively similar with those of the QOME, but the transient dynamics offer a basis for experimentally observable differences between the models. 

Further work to validate the accuracy of the thermometric QBME could involve comparing it to more exact numerical methods, such as the hierarchical equations of motion \cite{tanimura;1989,xu;2022}, to analyse its behaviour in different parameter regimes, and experimental testing of the predictions of the thermometric QBME. We have outlined a possible experiment, whereby a magnetic flux is rapidly applied to a shunted transmon and the system is allowed to decay, which can discriminate between the predictions of the thermometric QBME and the QOME at low temperature. The oscillations in the decay predicted by the thermometric QBME come precisely from its flux-translation invariant dissipation, and so experimental observation of this behaviour would not just serve to validate the model, but also to affirm the necessity of flux-translation invariance for accurate modelling of dissipative systems.

\begin{acknowledgments}
TMS acknowledges funding from the Australian Research Council Centre of Excellence for Engineered Quantum Systems (Project No.\ CE170100009). We thank Gerard Milburn for early conversations on this subject.
\end{acknowledgments}

\bibliographystyle{apsrev4-2}
\bibliography{reference.bib}

\appendix

\section{Steady State of Driven Transmon at Elevated Temperature \label{app:high T sweep}}

\begin{figure*}[!]
    \centering
    \includegraphics{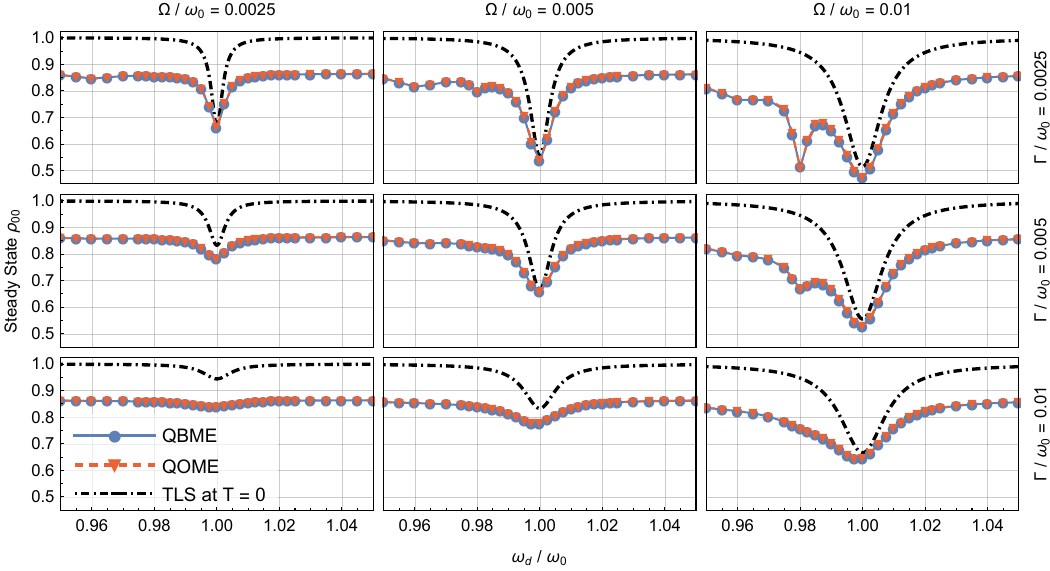}
    \caption{Each panel depicts the ground state population at equilibrium of a driven transmon shunted by a resistor (see \cref{eq:driven transmon ham}) as a function of the drive frequency. Parameter choices in each panel are the same as in \Cref{fig:rho00 low T sweep}, but at an intermediate temperature \(\kb T/\hbar \omega_0 = 0.5\). The ground state population is calculated with the thermometric QBME (circle, solid line), QOME (triangle, dashed line), and the analytical solution for a damped two-level system at zero temperature (dot-dashed line) (see \cref{eq:rho00 tls}).}
    \label{fig:rho00 high T sweep}
\end{figure*}

In the main text, we present the equilibrium ground state population of a driven resistively-shunted transmon as a function of drive detuning and for multiple values of the Rabi frequency and dissipation rate (\Cref{fig:rho00 low T sweep}). This was conducted at a representative temperature for superconducting device experiments (\(T = 15\mK\)), where the thermal energy is much less than the transition frequency. To further demonstrate the agreement between the thermometric QBME and QOME, we conduct the same sweep as in \Cref{fig:rho00 low T sweep} but at a temperature comparable to the transition frequency (\(\kb T/\hbar\omega_0 = 0.5\)). \Cref{fig:rho00 high T sweep} depicts these results and shows that the thermometric QBME and QOME agree very closely in the low to intermediate temperature regime as well.

\end{document}